\documentstyle[pra,aps,twocolumn]{revtex}
\def\p{\partial}

\input epsf

\begin{document}
\title{Instabilities of Higher-Order 
Parametric Solitons. Filamentation versus Coalescence.}
\author{Dmitry V. Skryabin 
\footnote{Electronic address: dmitry@phys.strath.ac.uk}, 
and William J. Firth}
\address{Department of
Physics and Applied Physics, John Anderson Building,\\ 
University
of Strathclyde, 107 Rottenrow, Glasgow, G4 0NG, Scotland, 
UK\\ URL: http://cnqo.phys.strath.ac.uk/$\sim$dmitry}

\date{December 23, 1997}

\maketitle

\begin{abstract}
We investigate stability and dynamics of higher-order solitary 
waves in quadratic media, which have a central peak 
and one or more surrounding rings. We show existence of 
two qualitatively different behaviours. For positive 
phase mismatch the rings break up into filaments which move
radially to initial ring. For sufficient negative
mismatches rings are found to coalesce with central peak, 
forming a single oscillating filament. 
\end{abstract}


Stability of optical solitary waves ('solitons') is one of the most
important questions of theoretical nonlinear optics because of its
direct connection with the possibility of experimental 
observation of solitons. Stability of the solitons in fully integrable
systems naturally follows from integrability. 
Solitons of the one-dimensional (1D) Nonlinear Schr\"oedinger 
Equation (NLS), describing propagation of short pulses in a 
fibre with cubic nonlinearity,
are a well known example \cite{ZakharovShabat}.  
A wide range of the non-integrable hamiltonian models also have
solitary solutions. For instance, equations describing
parametric interaction in quadratic nonlinear media have solitary solution, 
which were pioneered in Refs. \cite{Karamzin} and recently 
explored in details from both theoretical and experimental 
sides because of their many 
interesting features,  see e.g. \cite{Stegeman,KivsharPRL1}.  
In non-integrable systems the stability of ground-state solitary solutions 
is often governed by the derivative
of some integral invariant with respect to an associated free
parameter of the solution \cite{KivsharPRL1,Vakhitov,Pego}. 
For example, it has been rigorously proved that 
for ground-state bright solitary 
solutions of the generalized NLS equation positivity of 
the derivative of total energy with
respect to the nonlinear wave number shift is a necessary and 
sufficient condition for stability \cite{Vakhitov}. 
Numerical and analytic studies indicate that this 
holds also for ground states in quadratic media \cite{KivsharPRL1}.

Existence of higher-order solitary waves with bright and dark
central spots surrounded by one or more rings 
was demonstrated for 2D NLS equation with pure Kerr
\cite{HausYankauskas} and saturable 
\cite{Vakhitov,Soto91,Atai,prl97} nonlinearities
and also in quadratic nonlinear media  \cite{prl97,He,Torner}. 
No universal stability criterion is known for
higher-order bound states and their stability has to be studied 
individually in every case. It has been shown that for saturable  nonlinearity 
higher-order bound states with bright and dark central 
spots are stable with respect to purely radial perturbations,
obeying to criteria for ground states, but unstable with respect to
azimuthally dependent perturbations, showing breakup of their rings into 
filaments \cite{Soto91,Atai,prl97}. Properties of solutions with 
dark central spot are strongly affected by their nonzero angular momentum 
and these properties are very similar for both saturable and quadratic
nonlinearities \cite{prl97}. 

Dynamics induced by instability of higher-order states 
is fascinating phenomena on its own and it is a natural
starting point for understanding pattern forming phenomena in
the evolution of higher-order gaussian beams in nonlinear media 
\cite{DholakiaPRA}. Primary interest of this Communication
is to address (for the first time to our knowledge) problem of
stability and dynamics of higher order solutions with zero
angular momentum in quadratic media.
In particular we show that these solutions 
reveal new  scenario of evolution which is absent 
for corresponding solutions in Kerr-like media.
Namely for some parameters values symmetry-breaking instability
leading to filamentation along rings replaces with novel
symmetry-preserving instability resulting in coalescence
of rings with a central peak. 

We consider interaction of first and second harmonic optical 
fields propagating in a dielectric medium with quadratic nonlinearity,
under the conditions of type I phase matching and with 
negligible walk-off effects. The corresponding hamiltonian 
\cite{Karamzin}
is $H=\int\!\int dxdy~[\frac{1}{2}|\vec\nabla_{\perp}E_1|^2+
\frac{1}{4}|\vec\nabla_{\perp}E_2|^2 +\beta |E_2|^2-
\frac{1}{2}(E_1^2E_2^*+c.c.)]$, where
$\vec\nabla_{\perp}=\vec i\p_x+\vec j\p_y$ and $\beta$ is the
normalized phase mismatch.  All variables and parameters are 
dimensionless, and these scaled
units are used throughout the text and in the figures.
The evolution of the normalised
field envelopes of the fundamental ($E_1$) and second ($E_2$) 
harmonics obeys the system of equations: 
\begin{equation}
i\p_zE_m=\frac{\delta H}{\delta E_m^*},~~m=1,2. \label{eq1}
\end{equation}
We look for non-diffracting solutions of Eqs. (1) in the form
$E_m(z,x,y)=A_m(r)e^{im\kappa z}$, where 
$r={\sqrt{x^2+y^2}}$, $A_{m}$
are real functions and $\kappa$ is the nonlinear wave vector
shift.  The existence condition of localised
solutions with exponentially decaying tails is $\kappa >
max(0,-\beta /2)$. For any value of $\kappa$ in this range we
were able to numerically build higher-order many-ring solitary
solutions with a bright central spot.  Examples of spatial
profiles of
one- and two-ring solutions are presented in Fig. 1(a).  For any 
finite number of rings, the fundamental field has radial zeroes
but the second harmonic field always remains positive, though
having minima close to the zeros of the fundamental.  In the
limit $\beta\gg 1$ Eqs. (1) can be approximately reduced
to an NLS equation for the fundamental field. Accordingly
for increasing $\beta$ the second harmonic field tends to carry 
less and less of the energy.  The situation is
opposite for negative $\beta$, when $\kappa$ values are close to 
the boundary  of soliton existence.  Dependences vs $\kappa$ of the 
energy
invariant $Q=\int\!\int dxdy~(|E_1|^2+2|E_2|^2)$ are presented 
in Fig. 1(b).

To study stability we consider 2D perturbations of these 
solutions in the general form $E_m(r,\theta,z)=e^{im\kappa z}$
$(A_m(r) + \epsilon_m^{+}(r)e^{\lambda z+iJ\theta}+
{\epsilon_m^{-}}^*(r)e^{\lambda^{*} z-iJ\theta})$.  Here
$\theta$ is the polar angle and $J$ is the azimuthal
index of the perturbation. $J$ must be an integer for azimuthal 
periodicity. Linearising Eqs. (1) and putting
$\epsilon_{m}^{\pm}=r^{|J|}f_{Jm}^{\pm}(r)$ gives the
non-self-adjoint eigenvalue problem:
\begin{equation}
i\lambda_J\left[\begin{array}{c}
f_{J1}^+\\ f_{J1}^-\\f_{J2}^+\\f_{J2}^-\end{array}\right]=
\left[\begin{array}{cccc} 
\hat L_{J1} &   A_2       & A_1         &     0 \\
-A_2 &-\hat L_{J1} & 0 & -A_1\\ A_1 & 0 &\hat L_{J2} & 0 \\ 
0 &
-A_1 & 0 &-\hat L_{J2}\end{array}\right] 
\left[\begin{array}{c}
f_{J1}^+\\ f_{J1}^-\\f_{J2}^+\\f_{J2}^-\end{array}\right]
,\label{eq2}\end{equation} 
where $\hat L_{J1}=\frac{1}{2}\hat
R_J-\kappa$, $\hat L_{J2}=\frac{1}{4}\hat R_J-2\kappa-\beta$ 
and
$\hat R_J=\frac{d^2}{dr^2}+\frac{2|J|+1}{r}\frac{d}{dr}$.  
We
seek unstable eigenmodes in the class of functions obeying
$\frac{df_{Jm}^{\pm}}{dr}=0$ at $r=0$ and exponentially 
decaying at
$r\to\infty$. Corresponding eigenvalues belonging to the 
discrete
spectrum can lie anywhere in the complex plane outside the rays
$(i\Omega_c,i\infty)$ and $(-i\Omega_c,-i\infty)$  which
belong to the stable unbounded 
continuum, here $\Omega_c=min(\kappa,2\kappa+\beta)$. 
Unstable modes have eigenvalues with $Re\lambda_J > 0$. 
They must always 
have a counterpart with $Re\lambda_J < 0$ because of the 
hamiltonian nature of our problem.
Infinitesimal phase and translational transformations of the 
stationary solutions generate two  eigenfunctions:
$\tilde f_{0m}^{\pm}=\pm mA_{m}$ for $J=0$ and $\tilde
f_{1m}^{\pm}=\frac{1}{r}\frac{dA_{m}}{dr}$ for $J=1$, 
which are neutrally stable ($\lambda_{0,1}=0$).   

We will mainly deal with solutions with only one ring outside 
the central peak.  These show the main features of the dynamics 
of solutions with an arbitrary number of rings.
First we consider symmetry-preserving perturbations, $J=0$.
Using asymptotic techniques developed
for the ground-state solution \cite{KivsharPRL1}, it can be 
shown that the
neutrally stable mode branches at the point 
$\partial_{\kappa}Q=0$
giving instability for $\partial_{\kappa}Q<0$, see Fig. 1(b).
Thus we can conclude that the standard stability criterion for 
ground states \cite{KivsharPRL1,Vakhitov} is also a {\em necessary} 
condition for stability of higher-order bound-states. This instability is
related to the existence for $\partial_{\kappa} Q>0$ of a pair of
eigenmodes with purely imaginary eigenvalues (with opposite 
signs) lying
in the gap $(-i\Omega_c,i\Omega_c)$. At the 
point $\partial_{\kappa}Q=0$ these eigenmodes coincide with 
the neutral 
mode and for more negative $\beta$ appear again but with purely 
real eigenvalues of opposite sign, signifying instability. 
For the ground-state this is the only instability scenario and these
discrete eigenmodes disappear into the continuum for large 
$\beta>0$ \cite{EtrichPRE}. 

In the present system, we have undertaken numerical 
investigation of eigenvalue problem (2). The case $J=0$ reveals 
two pairs of discrete eigenmodes.  Interplay between them leads to a new 
bifurcation  scenario, which we study for different 
values of $\beta$ for fixed $\kappa=3$. 
Changing $\kappa$ at fixed $\beta$ has no qualitative effect due 
to scaling properties of (1).
However introducing this scaling modifies the stability criterion 
$\partial_{\kappa}Q>0$ \cite{KivsharPRL1}, which we prefer 
to avoid.

Real and imaginary parts of key eigenvalues from the discrete 
spectrum are
plotted vs  $\beta$ in Fig. 2.  In the limit of large $\beta$ we 
found one internal eigenmode (line 1 in Fig. 2) but at $\beta\simeq 
4.75$ another internal eigenmode 
(line 2) emerges from the continuum.  On emergence mode 2 has 
$Im\lambda_0=\kappa=3$, but as $\beta$ is decreased the 
eigenvalues of the two modes bring together, as Fig. 2 shows.  They 
fuse at $\beta\simeq -0.82$ to form two pairs of eigenfunctions 
with complex conjugate eigenvalues, giving onset of an 
instability (lines 3,4). At
$\beta\simeq -2.06$ a reverse bifurcation takes place (lines 5,6). 
One eigenmode (Fig. 3(a)) then has a purely imaginary 
eigenvalue (line 6) 
until it loses its stability at $\beta\simeq -5.61$ (line 7) where
$\partial_{\kappa}Q=0$. This is the standard instability scenario 
described above. The other eigenmode undergoes a similar bifurcation, 
but at $\beta\simeq -2.17$ (lines 5,8) which is well before the point 
$\partial_{\kappa}Q=0$. At this 
bifurcation the eigenfunction profiles, Fig. 3(b), are quite 
different from those of the neutrally stable eigenmode.  Thus, unlike the 
previously-known 
case (lines 6,7 and Fig. 3a) this new instability cannot be 
captured by asymptotic expansion around that neutral eigenmode.  
In both cases the unstable eigenvalues reach a maximum then go
steeply to zero near the existence limit of solitary solutions,
$\beta=-2\kappa$. 

The cascade of symmetry preserving bifurcations presented in 
Fig. 2 is somewhat similar to that for $TE_1$ mode instability 
in a planar waveguide (1D geometry) with Kerr nonlinearity 
\cite{Tran} where joint action of the refraction index discontinuties and field 
nodes leads to instability. In our situation 
the new symmetry-preserving instability develops in the region 
where the nodeless second harmonic starts to dominate over the 
fundamental, which has one or more nodes.

For symmetry-breaking perturbations, $J\ne 0$, the stability 
properties are not so rich as for $J=0$.  For $J=1$ our numerics reveals 
the presence of a neutral mode and a pair of eigenmodes in the 
discrete spectrum
with purely real eigenvalues, one of which is responsible for 
instability. We did not find any exchange of
stability of these modes. For every $J$ from $2$ to $5$ we find
such a pair of discrete eigenmodes with purely real
eigenvalues and all modes for $J>5$ belong to the continuum.
The unstable perturbations for $J=3,4$ are localised  around the 
ring of the bound-state in a manner similar to what happens 
in saturable media \cite{Soto91}.

To show how the character of instability of the one-ring solution 
depends on phase mismatch parameter we plot in Fig. 4 growth rates  vs 
$\beta$ for all unstable eigenmodes. For phase mismatches from the cascading 
limit down to $\beta\sim -3$ symmetry breaking
instabilities with $J=3,4$ are dominant. However, sufficiently far
from the NLS limit our new scenario, with azimuthally
homogeneous perturbations dominant, is realised. We stress 
again (see discussion above) that this symmetry preserving instability 
is not related to violation of the criterion $\partial_{\kappa}Q>0$.  
Note that in the limit $\beta\gg 1$ the $J=0$ internal eigenmode 
exists and the $J=1$ eigenmode has non-zero growth rate (Fig. 4(b)).

One expects the propagation dynamics of solitary states to be
mainly determined by the perturbation eigenmode with maximal 
growth rate. To examine this we performed an extensive series
of numerical simulations of Eqs. (1), using both polar and 
cartesian grids. 
Predictions based on our stability analysis are in good 
agreement with the results of our simulations.
An example of noise-stimulated break-up of a one-ring solution 
into three filaments is shown in Fig. 5(a).  We plot the real
part of the fundamental field profile rather than the intensity
distributions to show that the daughter solitons formed from the
ring are out of phase with the central one.  Radiation losses in 
the break-up are quite small, so that the initial energy $Q$ is 
mostly divided among the daughter solitons. 
 Their diameters are comparable to the width of the initial 
ring.  For $\beta$ values
where growth rates for $J=3$ and $J=4$ are almost equal the 
simulation results depended on the particular noise realisation, 
but we mostly observed the ring forming four filaments,
one of which was usually less intense than the others.  

Throughout the whole range of parameters
where the symmetry-breaking instability is dominant 
we observed repulsion between the central spot and daughter 
filaments, which results from the fact they are out of phase 
\cite{inter}, see Fig. 5(a). This repulsive force makes 
the outer filaments move out along radii (Fig. 5(b)),
in contrast to the tangential motion of daughter solitons after 
breakup of one-ring solitary waves carrying non-zero orbital 
angular momentum 
\cite{prl97}, where inter-soliton forces are negligible in 
comparison to the need to conserve angular momentum. 

Our stability analysis predicts a novel  symmetry-preserving 
instability scenario where the $J=0$ eigenmode 
dominates.  This prediction is indeed confirmed by the 
simulations. For example, at $\beta=-4.2$, instead of 
fragmentation we observed
coalescence of the ring with the central spot to form a single
filament. After transient dynamics this filament forms an
oscillating solitary wave, see Fig.6. These undamped pulsations 
are related to the existence of an internal eigenmode of the
ground-state solution \cite{EtrichPRE}. 

Considering now two-ring solitary solutions, we present the 
growth rates 
for the dominant eigenmodes and an example of symmetry-
breaking instability, see Fig.7.
General features of the dynamics are qualitatively similar to the
one-ring situation. 

The evolution of filaments following a symmetry breaking 
instability of peak-and-ring solitary solutions in saturable Kerr media 
\cite{Soto91} is a question which has not previously been examined. In 
simulations of this problem we observed the same sort of dynamics as described 
above for quadratic media, but with no coalescence phenomena.    

In summary, we have  undertaken a 
detailed analysis of stability of cylindrically-symmetric 
higher-order solitary waves due to parametric 
interaction in quadratic nonlinear media.  For a wide range of positive 
mismatches, symmetry-breaking instability
of the rings is predicted, and confirmed by simulations which 
show that the instability leads to filamentation into daughter 
solitons which are repelled radially from the central spot. 
For sufficiently negative phase mismatches we predict that a new symmetry-
preserving instability becomes dominant.  This is confirmed in 
simulations, in which the rings are found to coalesce with the 
central filament, forming an oscillating, single-peaked, solitary 
wave. To our knowledge this is the first explicit example of  a 
symmetry-preserving instability of 2D self-trapped beams in 
bulk media different from the Vakhitov-Kolokolov scenario  
\cite{Vakhitov}.
 
We thank G. K. Harkness, Y. Kivshar and A. Buryak for
discussions of relevant questions and L. Torner for sharing with
us  Ref. \cite{Torner} prior to publication. 
This work was partially supported by EPSRC grant GR/L 27916.

\begin{figure}
\centerline { \epsfxsize=8cm  \epsffile{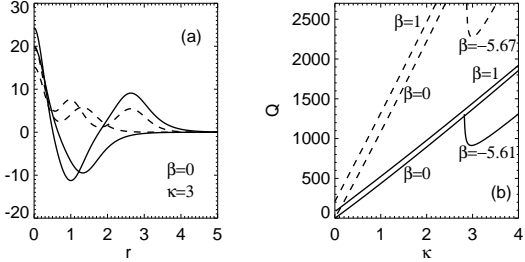}}
\caption{(a) Radial profiles of one- and two-ring solitary waves.
Full (dashed) lines are for $A_1$ $(A_2)$.
(b) Total energy vs $\kappa$ for one-ring (full lines) and two-
ring (dashed lines) solitary waves. The negative values of 
$\beta$ are chosen so as to give 
$\partial_{\kappa} Q=0$ at $\kappa=3$.}
\end{figure}

\begin{figure}
\centerline { \epsfxsize=8cm  \epsffile{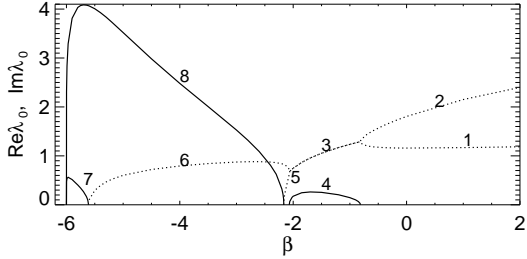}}
\caption{Real (full lines) and imaginary (dotted lines) parts of 
the eigenvalues of $J=0$ eigenmodes vs $\beta$, $\kappa=3$.}
\end{figure}

\begin{figure}
\centerline { \epsfxsize=8cm  \epsffile{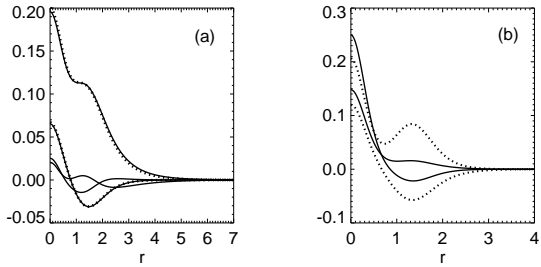}}
\caption{(a) The internal eigenfunction corresponding to branch 
6 in Fig. 2 at the point $\beta=-5.58$, slightly before the 
bifurcation point $\partial_{\kappa} Q=0$. 
(b) The internal eigenfunction corresponding to the bifurcation 
point at  $\beta=-2.17$ where branches 5,8 in Fig. 2 meet. Dots 
mark the neutrally stable eigenmode. $\kappa=3$.}
\end{figure}

\begin{figure}
\centerline { \epsfxsize=8cm  \epsffile{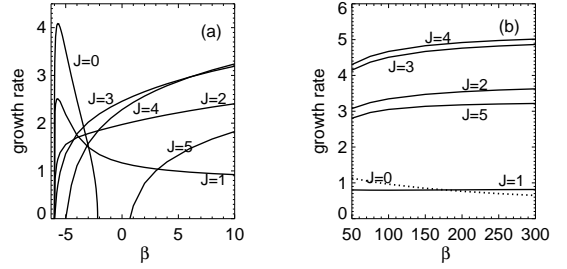}}
\caption{Growth rates of the maximally unstable eigenmodes vs 
$\beta$ for one-ring solitary solutions.
Dotted line in (b) displays $Im\lambda_0$ of
the $J=0$ internal eigenmode marked by Line 1 in Fig. 2. 
$\kappa=3$.}
\end{figure}

\begin{figure}
\centerline { \epsfxsize=8cm  \epsffile{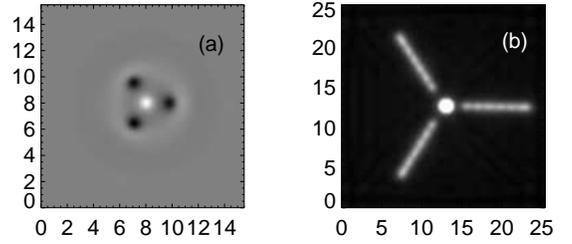}}
\caption{(a) Real part of the fundamental harmonic field at a late 
stage of a simulation of the symmetry-breaking process. 
(b) Superimposed images of its transverse intensity distribution 
at different values of $z$,  showing radial trajectories of the
daughter solitons. $\beta=-1$, $\kappa=3$.
Brightness and size  of central spot in Fig. (b) 
are exaggerated by the superposition of multiple images.}
\end{figure}

\begin{figure}
\centerline { \epsfxsize=8cm  \epsffile{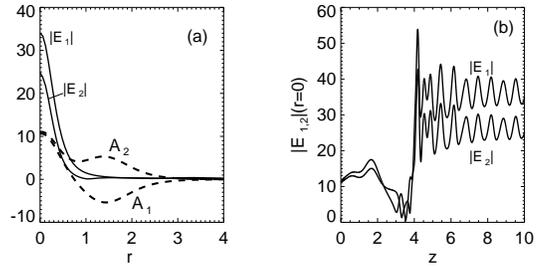}}
\caption{(a) Initial (dashed lines) and post-coalescence (solid 
lines) 
radial profiles of fundamental and second harmonics simulated 
where the symmetry-preserving instability is predicted to 
dominate.
(b) Corresponding evolution of $|E_{1,2}|$ at $r=0$ vs 
propagation coordinate $z$. $\kappa=3$, $\beta=-4.2$}
\end{figure}

\begin{figure}
\centerline { \epsfxsize=8cm  \epsffile{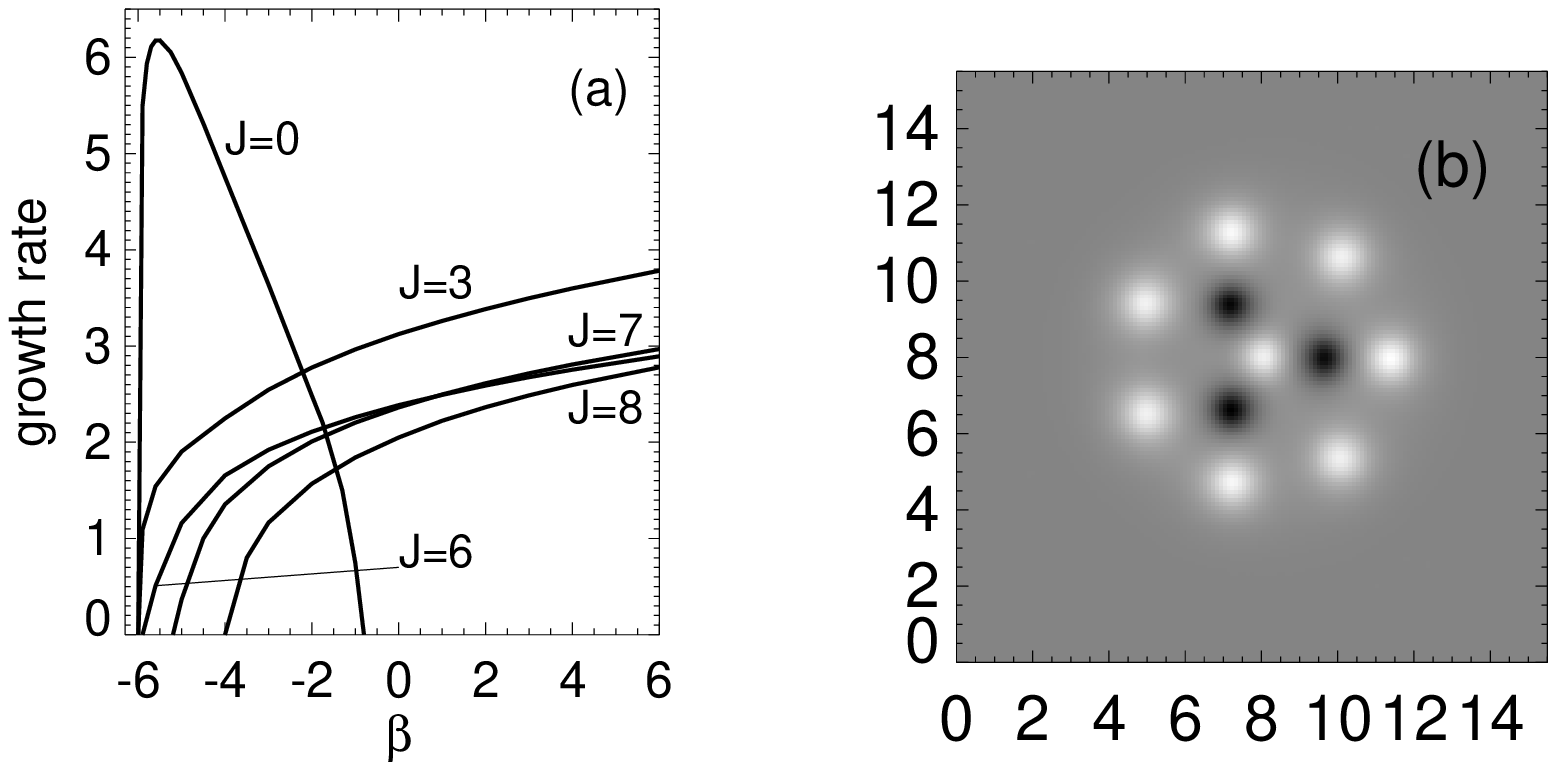}}
\caption{(a) Growth rates of the dominant unstable eigenmodes 
vs $\beta$ for two-ring 
solitary solutions. (b) Real part of the fundamental field at a 
late stage of a simulation of the symmetry-breaking process.  
$\beta=1$, $\kappa=3$.}
\end{figure}


\begin{references}

\bibitem{ZakharovShabat}
V.E. Zakharov and A.B. Shabat, Sov. Phys. JETP,  {\bf 34}, 
62 (1972).

\bibitem{Karamzin}
Y.N. Karamzin and A.P. Sukhorukov, JETP Lett. {\bf 20}, 339 (1974);
Sov. Phys. JETP {\bf 41}, 414 (1976).

\bibitem{Stegeman}
G.I. Stegeman, D.J. Hagan, and L. Torner, Opt.
Quantum Electron.  {\bf 28}, 1691 (1996) and refs. therein.

\bibitem{KivsharPRL1}
D.E. Pelinovsky, A.V. Buryak, and Y.S. Kivshar, Phys. Rev. 
Lett. {\bf 75}, 591 (1995).

\bibitem{Vakhitov}
M.G. Vakhitov and A.A. Kolokolov, Sov. Radiophys. {\bf 16}, 783 (1973).


\bibitem{Pego}
F.V. Kusmartsev, Phys. Rep. {\bf 183}, 1 (1989);
R.L. Pego and M.I. Weinstein, Phil. Trans. R. Soc. Lond. A
{\bf 340}, 47 (1992).


\bibitem{HausYankauskas}
H.A. Haus, Appl. Phys. Lett.  {\bf 8}, 128 (1966);
Z.K. Yankauskas, Sov. Radiophys. {\bf 9}, 261 (1966).

\bibitem{Soto91}
J.M. Soto-Crespo, D.R. Heatley, E.M. Wright, and N.N. Akhmediev, 
Phys. Rev. A {\bf 44}, 636 (1991).

\bibitem{Atai}
V.I. Kruglov, Y.A. Logvin, and V.M. Volkov, J. Mod. Opt.  {\bf 39},
2277 (1992); J. Atai, Y. Chen, and J.M. Soto-Crespo, 
Phys. Rev. A {\bf 49}, R3170 (1994).


\bibitem{prl97}
W.J. Firth and D.V. Skryabin, Phys. Rev. Lett. {\bf 79}, 2450 (1997).

\bibitem{He}
H. He, M.J. Werner, and P.D. Drummond, Phys. Rev. E {\bf 54}, 896 (1996).

\bibitem{Torner}
J.P. Torres, J.M. Soto-Crespo, L. Torner, and D.V. Petrov, 
J. Opt. Soc. Am. B {\bf 15}, 625 (1998).

\bibitem{DholakiaPRA}
V. Tikhonenko, J. Christou, and B. Luther-Davies, J. Opt. Soc.
Am. B {\bf 12}, 2046 (1995); J. Courtial, K. Dholakia, L. Allen, and M.J.
Padgett, Phys. Rev. A {\bf 56}, 4193 (1997);
J.P. Torner and D.V. Petrov, J. Opt. Soc. Am. B {\bf 14}, 2017 (1997);
P. Agin and G.I. Stegeman, J. Opt. Soc. Am. B {\bf 14}, 3162 (1997).

\bibitem{EtrichPRE}
C. Etrich, U. Peshel, F. Lederer, B.A. Malomed, and Y.A. Kivshar, 
Phys. Rev. E {\bf 54}, 4321 (1996).


\bibitem{Tran}
H.T. Tran, J.D. Mitchell, N.N. Akhmediev, and A. Ankiewich, 
Opt. Commun. {\bf 93}, 227 (1992).

\bibitem{inter}
Y. Baek, R. Schiek, G.I. Stegeman, I. Baumann, and W. Sohler,
Opt. Lett. {\bf 22}, 1550 (1997);
V. Steblina, A. Buryak, and Y.S. Kivshar, Opt. Lett. {\bf 23}, 156 (1998).

\end{references}
\end{document}